\documentclass[10pt,twoside]{epnt1p}

\usepackage{graphicx}
\usepackage{amssymb}
\usepackage{amsmath}
\usepackage{url}
\usepackage{mcite}

\begin{document}
{\hfill\normalsize\tt DESY 06-203}
\begin{frontmatter}

\title{Exotic Neutrino Interactions in Cosmic Rays}

\author[address1]{Markus~Ahlers}

\ead{markus.ahlers@desy.de}

\address[address1]{Deutsches Elektronen-Synchrotron DESY, Notkestra\ss e  85, 22607 Hamburg, Germany}

\begin{abstract}
The spectrum of extra-galactic cosmic rays (CRs) is expected to follow the Greisen-Zatsepin-Kuzmin (GZK) cutoff at about $5\times10^{10}$~GeV which results from energy losses of charged nuclei in the cosmic microwave background. So far the confrontation of this feature with CR data is inconclusive. In the absence of close-by sources a power-law continuation of the spectrum might signal the contribution of new physics. We have investigated the statistical significance of a model where exotic interactions of cosmogenic neutrinos are the origin of super-GZK events. A strong neutrino-nucleon interaction is favored by CR data, even if we account for a systematic shift in energy calibration.
\end{abstract}

\end{frontmatter}
\section{Introduction}
The appearance of extremely high energetic (EHE) cosmic rays (CRs) is a mystery. Simple geometric arguments show that the energy of nuclei originating in cosmic accelerators should be limited corresponding to the size and magnetic field of the accelerating environment (Hillas criterion)~\cite{Hillas:1985is}. Additional constraints arise from energy losses in the source that has to be balanced by the acceleration rate~\cite{Protheroe:2004rt}. In particular, the energy loss length of synchrotron radiation decreases with the third power of the magnetic field (at constant gyro-radius) and limits the efficiency of small candidate acceleration sites. The very few sources that seem to be capable of accelerating protons up to $10^{12}$~GeV include radio galaxy lobes and relativistically moving sources like jets of active galactic nuclei or gamma ray bursts.

Not only the acceleration mechanism of these particles seems to be problematic, but also their propagation in the interstellar medium. Resonant pion photoproduction in the cosmic microwave background (CMB) limits the range of EHE particles to a few $10$~Mpc. It was first emphasized by Greisen~\cite{Greisen:1966jv}, Zatsepin, and Kuzmin~\cite{Zatsepin:1966jv} (GZK) that this should be signaled by a cutoff in the CR spectrum at about $5\times10^{10}$~GeV if extra-galactic sources dominate at these energies. So far, the confrontation of this effective cutoff with CR data is inconclusive. In particular, the observations of two large exposure experiments AGASA~\cite{Takeda:2002at} and HiRes~\cite{Bird:1994wp,*Abbasi:2005ni,*Abbasi:2005bw} seem to imply conflicting results.

The ``excess'' of super-GZK events reported by the AGASA collaboration has led to speculations about a different origin of EHE CRs. Berezinsky and Zatsepin proposed that cosmogenic neutrinos~\cite{Beresinsky:1969qj,*Stecker:1978ah} produced in the decay of the GZK pions could explain these events assuming a rapid rise of the neutrino-nucleon interaction. We have followed this idea in Ref.~\cite{Ahlers:2005zy} and investigated the statistical significance of scenarios with strongly interacting neutrinos combining CR data from AGASA and HiRes and neutrino limits from horizontal events at AGASA and contained events at RICE.

The AGASA excess of super-GZK events could also be the result of a relative systematic error in energy calibration of $\pm30\%$~\cite{DeMarco:2003ig}. This interpretation is fueled by a recent re-analysis of the AGASA data~\cite{SHINOZAKI}. The preliminary results indicate that the energy of highly inclined showers has previously been over-estimated. If verified, this would result in a re-calibration of the spectrum according to a relative energy shift of $-10\%$ to $-15\%$. 

Remarkably, this energy shift agrees with predictions from a matching procedure of CR spectra assuming an early onset in extra-galactic proton dominance at about $10^9$~GeV~\cite{Aloisio:2006wv}. The calibration by the ``dip'' in these proton spectra resulting from electron-positron pair production on the CMB predicts a $-10\%$ and $20\%$ energy shift of the AGASA and HiRes data, respectively. We have incorporated this re-calibration into our analysis and show also the effect of the preliminary data from the Pierre Auger Observatory (PAO)~\cite{Sommers:2005vs} on our results.

\begin{figure*}[t]
\begin{minipage}[c]{\linewidth}
\centering\includegraphics[width=\linewidth]{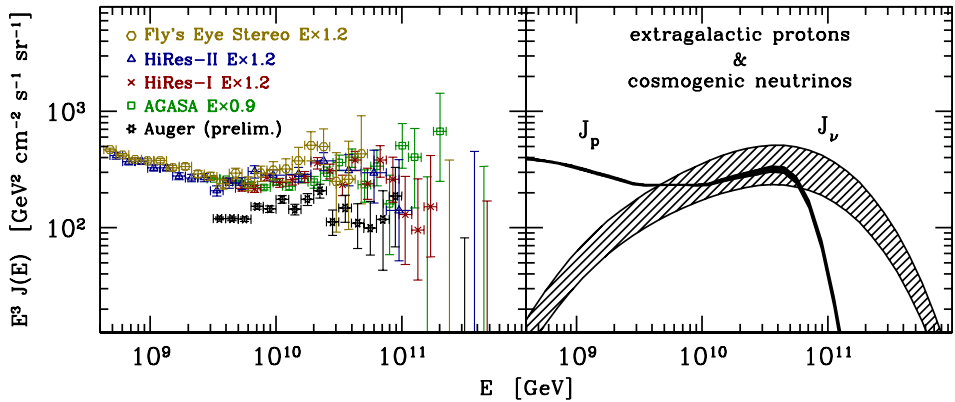}
\caption[]{{\rm Left panel:} The cosmic ray spectra from AGASA, Fly's Eye Stereo, Hires-I/II (with energy shift given in Ref.~\cite{Aloisio:2006wv}), and PAO. {\rm Right panel:} The flux of extra-galactic protons and cosmogenic neutrinos from CMB interactions corresponding to the 99\% CL of the fit.}
\label{fig1}
\end{minipage}
\end{figure*}

\section{Cosmic Sources of Protons and Neutrinos}

We assume that extra-galactic protons dominate the CR spectrum above $5\times10^8$~GeV. A convenient parameterization of a spatially homogeneous and isotropic source luminosity is given by
\begin{gather}\label{fluxp}
  \mathcal{L}_{\rm CR}({z,E}) \propto{(1+z)^n}\,{E^{-\gamma}\,e^{-\frac{E}{E_\mathrm{max}}}},\qquad
 z_{\rm min}<z<z_{\rm max},
\end{gather}
which accounts for an evolution with redshift $z$ and an exclusion of nearby ( $z_\mathrm{min}$) and early ($z_\mathrm{max}$) sources. In our analysis we have keep these parameters fixed at $z_\mathrm{min} = 0.012$, corresponding to $r_\mathrm{min} \approx 50\, \mathrm{Mpc}$, and $z_\mathrm{max} = 2.0$.  The power-law injection is effectively limited to energies below $E_\mathrm{max}$, which we also fix at $10^{12}$~GeV.

The flux of protons originating at distant sources is subject to energy redshift and collisions with the interstellar photon background during propagation. The dominating interactions of UHE protons are $e^+e^-$ pair production and meson photoproduction in the CMB (see e.g.~\cite{Engel:2001hd,*Fodor:2003ph,*Semikoz:2003wv,*DeMarco:2005kt,*Allard:2006mv}). These effects are taken into account by means of propagation functions, which have been provided by the authors of Ref.~\cite{Fodor:2000yi,*Fodor:2003bn}. The pions, which are resonantly produced at the GZK cut-off, decay into electron and muon neutrinos via the reaction chain $\pi^+\rightarrow \mu^+\nu_\mu \rightarrow\nu_\mu\bar \nu_\mu \nu_e\, e^+$ and the conjugate process. This constitutes a flux of UHE cosmogenic neutrinos which can be taken as a ``guaranteed source'' assuming extra-galactic protons as the highest energy CRs.

Depending on the environment of the proton accelerators neutrinos might also emerge in the decay of pions produced by  photoproduction processes during acceleration~\cite{Waxman:1998yy}. The flux of extra-galactic neutrinos have been estimated in Ref.~\cite{Ahlers:2005sn} assuming a low cross-over to extra-galactic protons at about $5\times10^8$~GeV. In contrast to a transition at the ``ankle'' this flux of neutrinos from the source dominate the total neutrino flux below $10^9$~GeV and exhausts current experimental neutrino bounds at energies of about $10^7$~GeV. The total neutrino flux beyond the GZK cutoff is however dominated by cosmogenic neutrinos and we will only use this contribution for our statistical evaluation. 

\section{Exotic Neutrino Interactions}

The flux of cosmogenic neutrinos associated with a low cross-over to extra-galactic protons is comparable to the flux of protons at the GZK cut-off as can be seen from Fig.~\ref{fig1}. A contribution of these neutrinos as super-GZK events requires a strong deviation from the feeble neutrino-nucleon interactions predicted by the Standard Model. In order to avoid large contributions in neutrino experiments the transition to a anomalously large cross section has to be very rapid. 
 
The realization of such a behavior has been proposed in scenarios beyond the (perturbative) Standard Model (SM), {\it e.g.} arising through compositeness, through electroweak sphalerons, through string excitations in theories with a low string and unification scale, through Kaluza-Klein modes from compactified extra dimensions, or through black hole and $p$-brane production, respectively. For details of these models we refer to some recent reviews, Refs.~\cite{Fodor:2004tr,*Anchordoqui:2005ey}. Skepticism about these scenarios has been raised in Ref.~\cite{Burdman:1997yg} based on limitations from s-wave unitarity or the naturalness of the couplings. However, these considerations do not necessarily apply for non-perturbative aspects like instantons.

In the following we will use a flexible parameterization of a strong neutrino-nucleon inelastic cross section ($\sigma_{\nu N}^{\rm new}$) focusing on three characteristic parameters: (i) the energy scale $E_{\rm th}$ of the new underlying physics, (ii) the amplification $\mathcal{A}$ compared to the SM predictions, and (iii) the width $\mathcal{B}$ of the transition between weak and strong interaction. A mathematical convenient parameterization is given by
\begin{equation}\label{cs}
  \log_{10}\left(\frac{\sigma_{\nu N}^{\rm new}}{\mathcal{A}\,\sigma_{\nu N}^{\rm SM}}\right)= \frac{1}{2}\left[1+\tanh\left(\log_\mathcal{B}\frac{E_\nu}{E_{\rm th}}\right)\right]\,\,.
\end{equation}
In general, experiments distinguish different CR primaries by their characteristic shower development in matter. For the sake of simplicity we will assume that the characteristics of the showers induced by strongly interacting neutrinos are indistinguishable from those induced by protons.  In particular, we assume for both primaries (i) a complete conversion of the incident energy into the shower, and (ii) equal detection efficiencies at the highest energies.

\begin{figure*}[t]
\begin{minipage}[c]{\linewidth}
\centering\includegraphics[width=\linewidth]{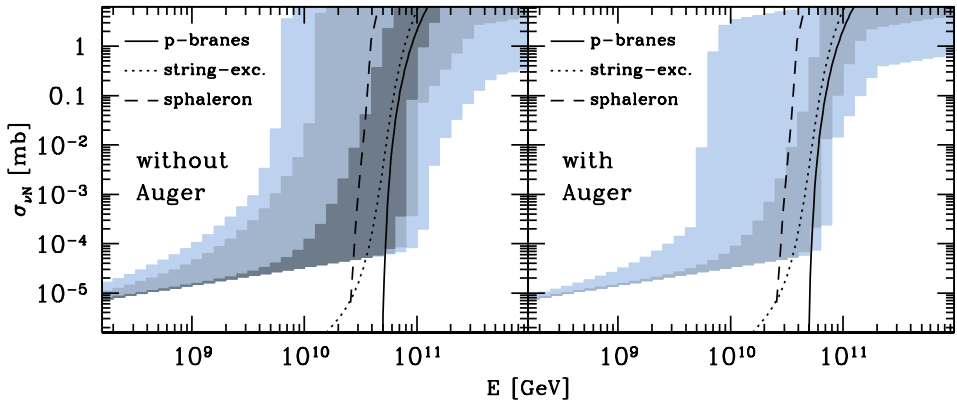}
\caption[]{The range of the cross section within the 95\%, 90\% and 68\% CL (outer to inner bands) obtained by the re-calibrated AGASA, Fly's Eye Stereo, and HiRes-I/II data, with (right panel) and without (left panel) the preliminary Auger data. Also shown are theoretical predictions of the neutrino-nucleon cross section enhanced by electroweak sphalerons, p-branes, and string excitations (see Ref.~\cite{Ahlers:2005zy} for details).}
\label{fig2}
\end{minipage}
\end{figure*}

\section{Quantitative Analysis}

Following the procedure of Ref.~\cite{Ahlers:2005zy} we have evaluated the CR spectra of AGASA, Fly's Eye Stereo, HiRes-I/II, and PAO (preliminary data) assuming extra-galactic proton dominance above and $5\times10^8$~GeV and a contribution of strongly interacting neutrinos as super-GZK events. In contrast to Ref.~\cite{Ahlers:2005zy} we have re-calibrated the energy by a multiplicative factor $0.9$ and $1.2$ for the AGASA and HiRes data, respectively, following Ref.~\cite{Aloisio:2006wv}. For consistency with neutrino experiments we have used the search results of horizontal events at AGASA and contained events at RICE~\cite{Kravchenko:2003tc}. Figure~\ref{fig2} shows the range of the cross section corresponding to the 68\%, 90\%, and 95\% confidence level (CL) from a goodness-of-fit test assuming cosmogenic neutrino fluxes from a low crossover scenario. The details of the statistical analysis and the approximations involved can be found in Ref.~\cite{Ahlers:2005zy}.

Compared to our previous results~\cite{Ahlers:2005zy} the re-calibrated spectrum is in a much better statistical agreement with the combined flux of protons and neutrinos from exotic interactions. This is mainly due to the fact that the calibration method ``optimizes'' the CR data to extra-galactic proton spectra below the GZK cutoff. However, within this particular model the re-calibrated data still favors an additional contribution from strongly interacting cosmogenic neutrinos. At the 95\% CL the corresponding neutrino-nucleon cross section should exhibit a steep increase by an amplification factor of $\mathcal{A}>10^3$. For $\mathcal{A}<10^5$ the transition should be very rapid ($\mathcal{B}<10$) at about $5\times10^{11}$~GeV. The allowed parameter space further shrinks if we also include the preliminary PAO data.

As an illustration of the result, we have considered three models of a rapidly increasing neutrino-nucleon cross section based on electroweak sphalerons~\cite{Han:2003ru}, $p$-branes~\cite{Anchordoqui:2002it} and string excitations~\cite{Burgett:2004ac}. The details can be found in Ref.~\cite{Ahlers:2005zy}. A separate fit of the source luminosity (Eq.~(\ref{fluxp})) with these cross section using the re-calibrated data gives a statistical acceptance at the 81\%, 65\%, and 60\% CL (without PAO), respectively. If we also include the PAO data in the fit, these values change to 98\%, 90\%, and 88\%, respectively, which is in reasonable good agreement with the results of Fig.~\ref{fig2}.

\section{Conclusions}

We have shown that in the absence of close-by sources exotic interactions of cosmogenic neutrinos might extend the cosmic ray spectrum beyond the GZK cut-off. Such a behavior is predicted in various extensions of the Standard Model, but might also be due to non-perturbative aspects like electro-weak sphalerons. In extension to our analysis of Ref.~\cite{Ahlers:2005zy} we have presented here the results of a fit to the AGASA and HiRes data, shifted according to the ``dip''-calibration from extra-galactic proton spectra~\cite{Aloisio:2006wv}. With these modifications a contribution of neutrinos is still favored by the data. The inclusion of preliminary data of the Pierre Auger Observatory weakens the results of the goodness-of-fit test.

\frenchspacing
\bibliography{refs}
\bibliographystyle{h-physrev3_mcite}

\setcounter{section}{0}
\setcounter{subsection}{0}
\setcounter{figure}{0}
\setcounter{table}{0}
\newpage
\end{document}